*Attilio Sacripanti*

# Astonishing Jūdō, first contact tactics

*A Biomechanical evaluation of tactical tools at start of high level competitions*

# Astonishing Jūdō, first contact tactics

## *A Biomechanical evaluation of tactics at start of high level competitions*

by *Attilio Sacripanti*\*†‡§

\*ENEA (National Agency for Environment Technological Innovation and Energy) Research Director
†University of Rome II "Tor Vergata", Italy
‡FIJLKAM Italian Judo Wrestling and Karate Federation
§European Judo Union Knowledge Commission Commissioner

**Abstract**
This paper is focalized on the limit application of judo throws, by tactics at first contact time, with some "astonishing" information at a first seeing, but biomechanically grounded, not often applied or because against the "sound common sense" or  out the old oral judo tradition.
To do so we provide an appraisal of the grips "concept" and his consequences in the Olympic sport "jūdō" from a biomechanics perspective, we will try to deeper  both the concept and the function of grips and define the potential application of some throws "without grips".
Broadening this situation we try to underline some specific throwing situation in which grips are or not at all applied or applied in non conventional way.
We describe at first the problem from the theoretical point of view.
And as second point we try to find practical application, original or already developed in high level competitions.
The provocative words "Judo without grips" or "throw without grips" are connected to the limit application of some biomechanical tricks, grounded on two well known physical principles: the time advance in the attack, in Japanese Sen no Sen (already applied in real competitions), and the utilization of the own inertia connected to high attack speed to apply in totally original way one of the two biomechanical tools utilized to throw the human body.

*Attilio Sacripanti*

# Astonishing Jūdō, first contact tactics

*A Biomechanical evaluation of tactical tools at start of high level competitions*

*Attilio Sacripanti*

# Astonishing Jūdō, first contact tactics

*A Biomechanical evaluation of tactical tools at start of high level competitions*

## 1. Introduction

Judo Olympic Sport every year more extended in the world, is in continue evolution during since his appearance in Olympic world in 1964.
The most advanced Judo Countries (France, Korea, Japan, and Nederland, Russia) try to find new way to improve their results every year.
In this dynamic situation, this paper try to analyze new applicative tools from high level competition match analysis, and propose some tricky ideas to find new astonishing way in Judo interaction (throw).
The field of study will be the transitory phase between two free bodies and two connected bodies during judo high level competition, this area totally lack of every serious technical study, applying biomechanics we will try to find solutions also out the common basic judo teaching tradition.
 To do so we provide an appraisal of the grips "concept" and his consequences in the Olympic sport "jūdō" from a biomechanics perspective, we will try to deeper  both the concept and the function of grips and define the potential application of some throws "without grips".
Broadening this situation we try to underline *some specific throwing situation in which grips as main tool to throws, are: or not at all applied, or applied in non conventional way.*
We describe at first the problem from the theoretical point of view, focalizing on the choose tools.
And as second point we try to find practical application, utilized in high level competitions.
The provocative words "Judo without grips" or "throw without grips" are connected to the limit application of some biomechanical tricks, grounded on two well known physical principles: the time advance in the attack, in Japanese (Sen no Sen), and the utilization of the own inertia connected to high attack speed to apply in totally original way one of the two biomechanical tools (couple and lever) useful to throw the human body.

## 2. The basic meaning of grip
   *(Kumi Kata classical vision)*

Normally in classical vision the coach or teacher starts from ki hon kumi kata (basic grip in judo as the Japanese tradition).

Ki hon kumi kata (組方) [1] are the fundamental grips: right hand at the left collar on a level with axilla, to seize judogi with little finger, annular, middle finger, leaving in relax thumb and index; left hand at the right sleeve on a level with elbow sizing a lot of textile. These grips are the most natural and simple

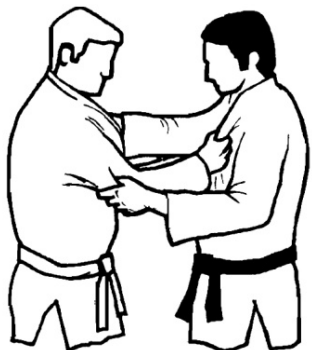

**Fig 1** *basic grip*

There are naturally a lot of variations depending on competitor's build, on style, on positions, on relative strength. It's important to remember that grip has to remain relaxed during the attack or defense in club. The apprentice must study this fundamental grip whether during static training with competitor (*uchi komi*) (打ち込み) or during dynamic phase (*kakari geiko*) (懸かり稽古); in fact, it's really important to learn the 'contact' with competitor using *kumi kata*.

The kinetic superior chains (arms) have, in the classical vision, four roles in the *kumi kata.:*

1) <u>**Active role: to transfer to the competitor's body an impulse to realize throwing technique.**</u>
2) <u>**Passive role: to stop the impetus and the movement of the competitor during his throwing technique.**</u>
3) <u>**Advising role: to receive information from the adversary's body about his movements**</u>
4) <u>**Alert role: to receive from the adversary body's movements alert about his attack action**</u>

In the ki hon kumi kata the right arm, in a passive way, takes information about directions of the moving body of competitor while the left arm takes the information about direction of his unbalance.

In active way, the right arm affects a motion of translation to the 'masse' of the competitor body, while the left arm brings out the moving action with an accurate directional character.

In a more advanced level is important to study the best possible way to change the grips depending on opportunity, with minimum energy.

Considering in this area only the role of superior chains the following figure shows the connection among different kind of grips on the adversary body.

In this classical optics many studies are focalized on the maximal strength of the arms that grip the adversary body. [2] Both isotonic and isometric [3], on the endurance of arms measured by lactate concentration [4] and also by the study of connection between strength of hand dominance and weight categories [5].

However other classic studies found not high difference in strength between judo athletes and non judo athletes but in judo athletes was found higher endurance.

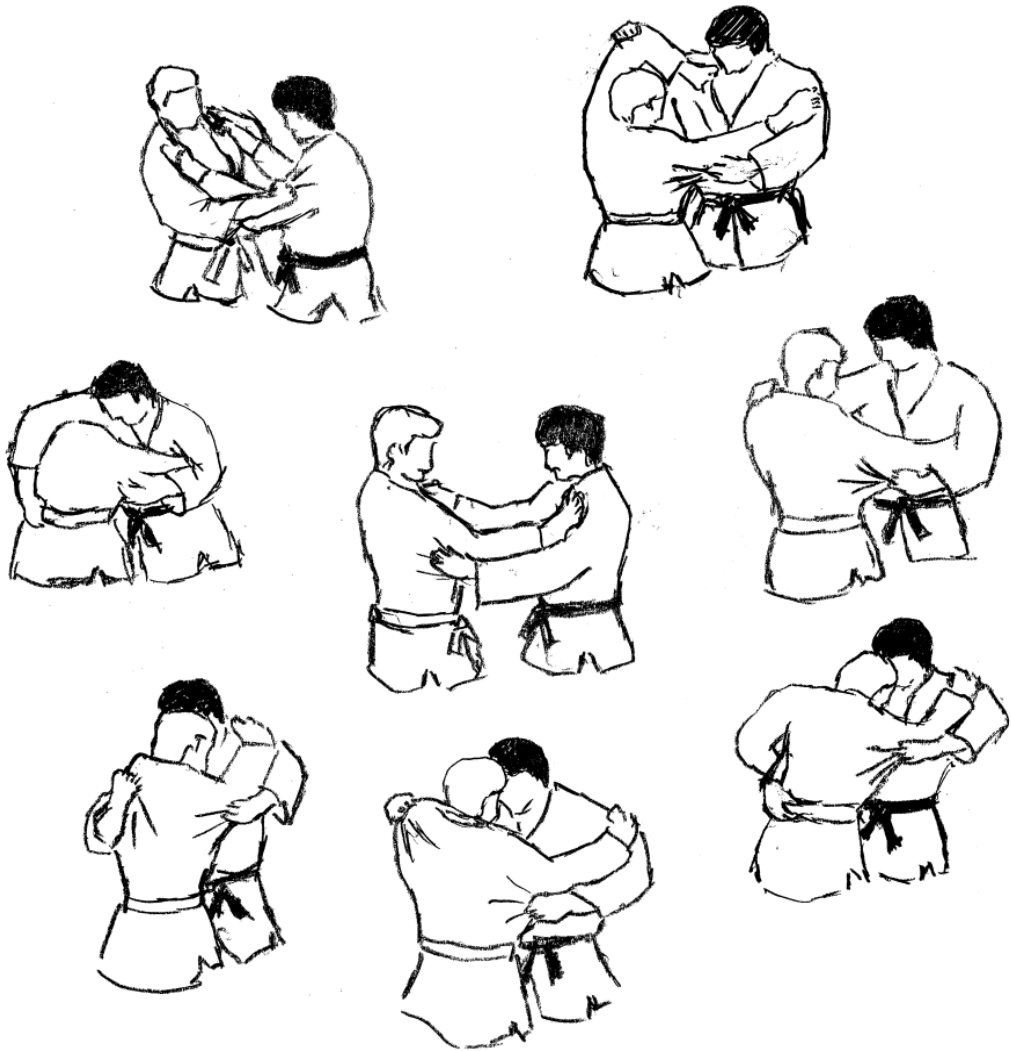

*Fig.2 Kumi Kata Superior Kinetic chain, various positions.*

At this level the kumi kata develops not only the capability *to sense information about rival movements,* but also the capability *to impose him own initiative physical and psychological*.

The search for a right energy utilization, takes probably a variation on the balance position of the competitor (with kumi kata, in general, increases muscular body tension) and in the weight relative distribution consequently, it rests on the ball-feet.

Some Japanese studies (*Studies on Judo techniques with respect to distribution of body weight*) [6] have proved that the weight distribution, in the tested competitors, was much more forward; you pass from natural and erect position *shizen hon tai* to the natural position with grip on a rival, during this action the muscular tone increases and the body movement becomes more dynamic, these positions ought to give the action very fast and a right start position in the attack.

The consideration of the dynamic balance in the athletes couple, like a single bio-kinematic group, proved that this couple is in permanent balance, through a whole of tensions, tractions and restraining reactions, even if each athlete has a position of abnormal unstable balance.

At the light of these facts, you can make a correct analysis of competition only if you take into account athletes, considered like an exclusive whole: the biodynamic grouping *"athletes couple" (couple of athletes).*

For this reason it is fundamental, realizing a technique, to use at best the concept of *relative distance*.

Better distance to realize a technique is in the *balance* between optimal **Time** to have a contact and the available **Space** to realize the attack movement, preserving a large **inertial momentum** in the athlete's couple. [6]

The relative distance between athletes in biodynamic grouping can be annulled to realize a technique by a *taisabaki* or controlled along longitudinal axis by *guard positions*.

During a competition there are many sorts of approaches which regulate relative distance between two athletes.

All these movements are called General Action Invariants, [7]. In order to didactics, you can group all the grips positions in three classes:

## *2.1 Guard Position*

### Normal guard positions

Two athletes are placed so that relative distance between them is more or less the same on both sides. In this kind of guard positions there is also the one called by Japanese SHIZEN HON TAI (自然本体)

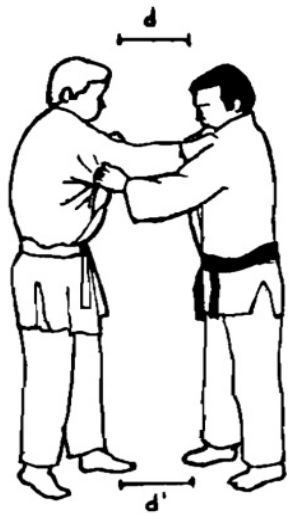

**Fig.3** *Normal guard d=d'*

### Diagonal guard positions

The athletes are placed so that relative distance between them is from one side very closed, while on the other side is large and sometimes open.

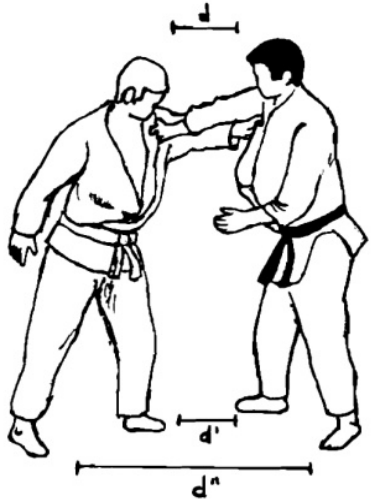

**Fig 4** *diagonal guard d=d'<d''*

### Curled up guard positions

Two athletes are placed so that relative distance between them at shoulder level is very close, while at hips level is larger possible, among these guard positions there is one called by Japanese JIGO TAI. (自護体).

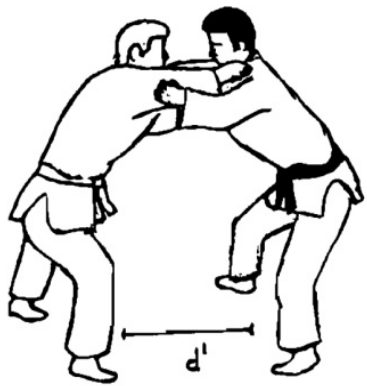

**Fig.5** *Curled up guard d<<d'*

To everyone is associated a special group of *Kumi Kata*, to each of these guard positions will coincide a particular strategy competition and the capability to realize some throwing techniques. [8,9] .

Other finding analyzed is the capability of athletes to apply throws at right and left side of the adversary, in order to increase the technical-psychological pressure on the adversary, some interesting data are shown in few papers, the following table is relative to young Spanish judo athletes [10]-

| Técnica | Índice de ambidextría (%) |
|---|---|
| seoi-nage | 44.4 % |
| sukui-nage | 33.8 % |
| ura-nage | 28.6 % |
| de-ashi-harai | 27.5 % |
| kata-guruma | 23.2 % |
| hiza-guruma | 21.8 % |
| uchi-mata | 20.4 % |
| osoto-gari | 17.6 % |
| kosoto-gake | 16.9 % |
| o-goshi | 10.7 % |
| tomoe-nage | 9.9 % |

*Tab. 1. Percentage of ambidextrous throws application in Young Spanish athletes.[10]*

In the dynamic field to these guard positions are connected often a well-defined cadence or rhythm of attack in the biodynamic group "athletes couple".

These guard position are closely connected to the well known "*Competition Invariants*" [6].

Nowadays, with high professional specialization at international level, the evolution of superior Judo is present above all in two well-determined biomechanical aspects:

1) **Increasing physical strength**
2) **The Kumi Kata predominance, which originate from this approach.**

From the study of world championship, Olympic games and of other important competitions and from *shiai* experience, points out some international competitors became a specialist about particular *Kumi Kata* to be able to effect their special technique (*Tokui waza*) in dynamic conditions.

From the study of initial phase of high level competitions you can comprehend the match crucial moment is "*the fight to impose one's Kumi Kata*"; many times, before *Kumi Kata,* there are some "*hand jumps*" which stabilize some *safety Kumi Kata* and permit to pose himself in profile towards adversary; in this way, the rival hasn't a great surface to grip and to attack, and so he's able to avoid the eventually believed dangerous and unfavorable grip (Diagonal guard positions).

After this phase, athletes try to build carefully and slowly their favorite position of *Kumi Kata.*

Sometime a little absent-mindedness and a brief weakness, during control of points of contact (hands) moving, becomes a decisive chance to win a match for a technically well-prepared adversary.

Nowadays, the realization and definition of a special *kumi kata*, during a competition, is considered an art in itself and a fundamental phase of *"Dynamic Superior Judo"*.

In the previous dynamic analysis we have already talked about important role which hands, as points of contact, play; therefore, only by an arms action, with a right body moving, will be possible to unbalance one of the athletes in biodynamic grouping "athletes' couple".

Many times, today, the defeat is much more in a *kumi kata* error; because *kumi kata* errors are directly reflected in the balance (whole body position) and in the energy transfer to the single athlete.

During "high level dynamic Judo" practice, the *kumi kata* hasn't only an informative role but, above all, it becomes an essential psychological aspect about imposition of one's will to the adversary to cross his reactions.

All these information produce some fundamental body's characteristics required for every athlete that inform both their conditioning and competition training:

--*Increasing physical strength*
--*Increasing acrobatics capabilities*
--*During a match high and constant "rhythm"*
--*Capacity to use one's special kumi kata, after to have had the adversary kumi kata*
--*To study the opportunities realizing one's special kumi kata*
-- *Continual adjournment about Renzoku waza*（連続技）*produced by one's special kumi kata*

## 2.2  Grips and Their Classical Objectives

The classical objectives of gripping are two

1. place one hand (Tsurite) (釣手)  in position to push Uke back toward the mat
2. Secure a locking hand (Hikite) (引手) that will prevent the defender from getting away from your throwing action

The tsurite hand cannot be placed without preparation. Place it by degree grabbing the nearest part of opponent's judogi, sleeve, elbow, or lapel; then grasping the adversary's judogi push down.

When pushing down keep your gripping hand tucked in close to your side and move your body (gripping side) lowering it under your grip.

The reaction to the push will expose part of your adversary's judogi to your grips, then, in such way, it is possible to improve your grip position

**Hikite – Tsurite**
Gripping methods fits into two categories within jacket work:
1 Hikite (the main pull) Sleeve grip "the working hand" or "Long pull"
2 Tsurite (the drawing hand) Lapel grip "the playing hand" or "short pull".

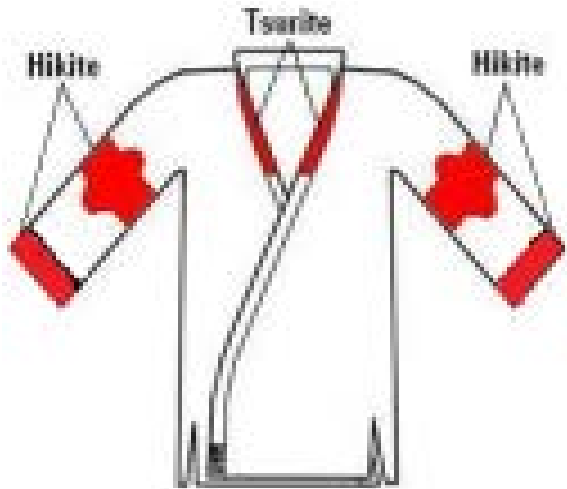

*Fig6 Specific Judogi Areas (on the edge it is a non classical Hikite grip)*

A few comments are in order regarding the definitions of hikite and tsurite.

**HIKITE**
The pulling hand is utilized to transmit tension to uke's shoulder joint. The hikite pull is a large amplitude pull, that normally account of 60% of the forward rotational unbalance. Properly done by Tori, it results in the desired upper shoulder girdle rotation of the Uke, when combined with the action of the tsurite hand.
Normally this action could be developed in two dimensions:

1. Vertical dimension given by position of grip in on location between elbow and shoulder
2. Horizontal dimension of the grips on the arm ( internal Vs external; upper Vs lower )

The Hikite traction has the goal to open the angle between homers and trunk of Uke.
If the grip is placed just above the wrist area when the pulling is applied, the Uke's arm must first essentially lock straight at the elbow before the shoulder joint senses the fully applied tension necessary for the shoulder girdle rotation.
Too high a grip on Uke's arm is undesirable since this betters the amplitude but worsen the mobility of the Tori arm.

Generally, optimum arm pull on uke's arm should be initially horizontal and parallel to the mat surface. Contrary to popular belief, raising the Uke's arm above the horizontal, it limits the effectiveness of the unbalance action; the only reason that this habit was introduced in judo praxis is that in this way uke's body stiffens himself like a rigid body losing his avoidance capability.
 It is important to remember that the overall pulling is accomplished and effective with Tori using his/her upper body in conjunction with the pulling hand.
This is usually rotation or translation of the trunk.
**Special Use**: the Hikite arm can be also utilized in a complete different way pulling down the adversary arm along his body, in order to apply the maximum pressure on one adversary's side (Full Rotational Application [9]).

## TSURITE
Tsurite is the pulling force action to lapel of judogi, normally called short amplitude pull, it could account for the 40% of forward unbalance.
Tsurite comes from the word tsuri meaning to fish and incorporates the idea of drawing an opponent off balance, just as a fishing rod bends when it draws a fish out of the water.
It is very important for the right application of Tsurite movement that arm turns at elbow and shoulder level.
A grip near the head helps mobility for unbalance but needs more force and gives less mobility for arms and legs.
Inversely a low grip gives less amplitude for unbalance, but also less force and helps arms and legs mobility.
The strongest position of the lifting hand is when the elbow of the Tori is placed in contact with the upper body with the hand located directly above it.
The weakest position of the Tsurite is when the arm is fully outstretched.
 The hand has zero lifting capability in this position.
Experienced athletes in competition usually made grip change that afford advantage than a disadvantage in throwing capability.
Most athletes usually vary the Tsurite grip far more often than the Hikite grip during tournament play.
**Special Use**: the Tsurite arm can be also utilized in a complete different way, blocking and pulling down the adversary shoulder or one his body's point along, in order to apply the maximum pressure on this point blocking the adversary's side (Full Rotational Application of Throws [9]).
Gripping is, in his classical vision, the first line of attack and defense. A great deal of time in a match is now devoted to securing your grip the so called grip fighting.
Because once high level athletes have the desired grip, they are able to control the match.

***In our vision "the First Contact" before the stabilized desired grip is the key time and a wide area to apply original or new concept in judo fight***.

Many fighters will attack quickly as soon as they have a grip some others have developed the ability to attack with only one hand.
High level athletes are well advised to prepare for fighting against these opponents by acquiring the ability to fight in the same way (diagonal *guard position*).
However it is well known that in standard judo view grip (arms) positions are connected to throws, this view thanks to the biomechanical analysis changes in guard positions ( whole body) that link and connect both preferred pace and preferred throws.
Studies on throwing techniques percentage of usage and tactics related are well known in judo science,
 In table 2, for example, it is possible to see effectiveness of throwing actions in London Olympic.[11]
Instead table 3 shows some result relative to the World, European championships and Olympic Tournament between 2010 -2012. [12]
Following the author "Combat model structure consists of technical and tactical procedures, mainly  shoulder and leg structure occupying more than 65% of all action taken."
Mainly this study was focalized on the way to build a technical-tactical action.
The author affirms that his model is "significantly different  from previous models presented in the literature" because the percentage and procedures were different.

| *Throws Effectiveness In London  Olympic 2012* | | |
|---|---|---|
| *Throws* | *Effectiveness  Male  %* | *Effectiveness  Female  %* |
| Seoi  ( Ippon – Morote - Eri) | 14.8     (329) | 8.2    (222) |
| Uchi Mata | 9.2     (138) | 15     (143) |
| O Uchi Gari | 15     (53) | 24     (49) |
| Ko Uchi Gari | 12     (57) | 37     (35) |
| Tai Otoshi | 25     (36) | 23.8     (21) |
| Soto Makikomi | 10     (10) | 23.6     (17) |
| Tani Otoshi | 46     (13) | 50     (16) |
| Uchi Mata sukashi | 90     (10) | 100     (10) |
| *Couple* | *28.7* | *39* |
| *Lever* | *24* | *26.4* |

*Tab.2 Male and Female Throws Effectiveness in London*

| No | The name of the technique, according to the international classification | The total number of attack | Actions performed on the left side | Actions performed on the right side | percentage of shares (with score) |
|---|---|---|---|---|---|
| 1 | SEOI-NAGE | 216 | 129 | 87 | 20 |
| 2 | UCHI- MATA | 170 | 102 | 68 | 16 |
| 3 | TAI-OTOSHI | 40 | 22 | 18 | 3.7 |
| 4 | O-UCHI-GARI | 74 | 44 | 30 | 7 |
| 5 | KO-UCHI-GARI | 48 | 37 | 11 | 4.4 |
| 6 | HARAI-GOSHI | 47 | 25 | 22 | 4 |
| 7 | O-SOTO-GARI | 32 | 18 | 14 | 3 |
| 8 | KO-SOTO-GARI | 30 | 17 | 13 | 2.77 |
| 9 | YOKO-OTOSHI | 21 | 12 | 9 | 2 |
| 10 | SEOI-OTOSHI | 26 | 14 | 12 | 2.4 |
| 11 | TANI-OTOSHI | 23 | 13 | 10 | 2.1 |
| 12 | KATA-GURUMA | 31 | 17 | 14 | 2.87 |
| 13 | DE-ASHI-HARAI | 32 | 17 | 15 | 3 |
| 14 | SASAE-TSURI | 8 | 3 | 5 | 0.74 |
| 15 | MOROTE-GARI | 30 | 16 | 14 | 2.77 |
| 16 | TOMOE-NAGE | 43 | 25 | 18 | 4 |
| 17 | SUMI-GAESHI | 35 | 15 | 20 | 3.24 |
| 18 | O-GOSHI | 41 | 24 | 17 | 2.87 |
| 19 | URA-NAGE | 14 | 10 | 4 | 2 |
| 20 | SUMI-OTOSHI | 20 | 5 | 15 | 2 |
| 21 | KIBISU-GAESHI | 18 | 9 | 9 | 1.66 |
| 22 | USHIRO-GOSHI | 30 | 17 | 13 | 2.77 |
| 23 | SODE-TSURI-KOMI- | 25 | 15 | 10 | 2.31 |
| 24 | OTHER | 26 | 16 | 10 | 2.4 |
| 25 | TOTAL | 1080 | 622 | 458 | 100 |

*Table 3. Competitive model to technical and tactical actions conducted in the fight standing [12]*

Today acquiring right grip position in high level competitions is so much important that grip fighting has become an boring attitude prevalent in competition where the fight for grip has resulted in the development of overly defensive tactics.

## *3. A short survey of Researches on competitions*

This attitude in real contests, focalized also a lot of advanced researches on the connection between grips and throws in different condition or with different adversaries [13, 14, 15, 16] or grips and tactics [17, 18] .
Other field of interest not strictly connected to grips but connected to tactics in competition, was the study of utilization of time during contest, these studies were connected to tactics of attack or pace of attack in high level competition [19, 20].
Tactics in competition is a main argument for many studies, starting from the historic work of Sterkowicz and Maslej on the throw tactics in high level competitions [21] not the first but very complete, till to the recent work of Miarka, Calmet and Boscolo del Vecchio [22] a complete and analytical review of techniques and tactics in judo, from 1996 to 2008.
In these studies tactic analysis is grounded on a comparisons of frequency distribution of events over a range of factors such as: time structure, number of applied techniques and directions, number of successful actions per minute, quality of attacks awarded with points, movements of elite judo athletes development, and grip forms.
In the next table there are shown the grip form percentage in high level competition referring to a work of Werrs 1997.

| Group  | Same Grip | Opposite | Sleeve | No Grip |
|--------|-----------|----------|--------|---------|
| Male   | 8         | 45       | 4      | 43      |
| Female | 14        | 50       | 6      | 30      |
| Mean   | 10        | 48       | 5      | 37      |

**Tabla 4: Grip configurations (%) Male and Female athletes A t l a n t a   O l y m p i c 1996.[ 2 2 ]**

In these times researchers look more interested in connection among grips configuration and throws applied.
It is well known that in judo, two athletes get close to each other, perform their grasps, and move on the mat and attack.
The approaches and kumi-kata type (grip form) give way to specific behaviors between the two contestants, but again we specify that it is the whole body position into the Couple of athletes system that drives the throw effectiveness..
Some studies by Courel and co-workers [23] tray to identify the effects of grip laterality and throwing side combinations (i.e., attacking on the same side of the gripping, or vice versa) on attack effectiveness and combat result in elite judo athletes.
The results were that attacking on the same side of the kumi-kata increase the chance of scoring and winning the combat independently of sex and weight category. Perform same-side attacks by kenka-yotsu (adversaries using reverse grip, right versus left) it was the most effective one, especially for lightest weight judo fighters.
 Both perform same-side attacks by ai-yotsu using right or left grip at the same time; and only one athlete gripping (only the athlete attacking performing the grip);are situations more open and increases the likelihood of winning the combat for both opponents.

One other recent study [24] compares the throwing techniques efficiency index performed from the same and opposite grip for senior male and female during Bosnia and Herzegovina senior State championships. Male dominate in regards to the same side grip, while female dominate in regards to the opposite side grip. The most efficient throwing technique for male considering the same side grip was as expected Ippon seoi nage, while for the female Harai goshi.

The highest efficiency technical index both for male and female regarding the opposite side grip was for Uchi mata

Some new data about connection between grip configurations and throwing techniques are presented in [25] from which we report the following table

| MALE | | | | FEMALE | | |
|---|---|---|---|---|---|---|
| Same grip | Opposite grip | Sleeve and grip | Rank | Same grip | Opposite grip | Sleeve and grip |
| SUK 11.3 | SUK 6.0 | ISN 6.4 | 1 | SUK 6.0 | OUG 6.3 | ISN 5.4 |
| UMA 6.0 | TNO 6.0 | STO 1.7 | 2 | OSG 5.9 | TNO 5.7 | SMK 2.9 |
| OUG 4.0 | KGU 4.5 | SMK 1.1 | 3 | OUG 4.3 | UMA 5.4 | STO 2.0 |
| KGU 4.0 | SUG 3.4 | KUG 0.4 | 4 | TNO 4.3 | SUK 4.0 | OUG 0.6 |
| SON 3.8 | UMA 3.2 | | 5 | UMA 4.0 | OSG 2.9 | |
| TNG 3.2 | ISN 1.9 | | 6 | ISN 3.7 | DAB 2.9 | |
| ISN 2.6 | OUG 1.9 | | 7 | KGU 3.7 | HRG 2.6 | |
| SUG 2.6 | DAB 1.9 | | 8 | KUG 2.9 | KGU 2.3 | |
| KUG 2.3 | KUG 1.5 | | 9 | SON 2.3 | ISN 2.0 | |

*Table 5. The most frequent throwing techniques using different gripping configurations between male and female cadets in percent values.[25]*

***International Judo Federation (IJF) codes of throwing techniques:*** SUK – *sukui nage*; UMA – *uchi mata*; OUG – *o uchi gari*; KGU – *kata guruma*; SON – *seoi nage*; TNG – *tomoe nage*; ISN – *ippon seoi nage*; SUG – *sumi gaeshi*; KUG – *ko uchi gari*; TNO – *tani otoshi*; DAB – *de ashi barai*; STO – *sode tsurikomi goshi*; SMK – *soto maki komi*; OSG – *o soto gari*; HRG – *harai goshi*.

Only two studies in these years are connected to the study of first contact situation regarding Kumi Kata [26, 27] and some older one's refers to are paper of Weer 1997, but no one was focalized on the identification of the physical-biomechanical tools utilized by Tori during this high dynamic phase of the contest, to grasp and throw the adversary, enhancing attack effectiveness .

## 4. A deeper meaning of grip

**A well known sentence give the most classical information about grips**: … *without a grip in Judo you will not be able to throw. This is easily demonstrated, try and throw someone with no grip at all ( non touching), then with one hand on your opponent, then with two hands….*
This very sound "Common Sense" give us the right direction to truth…. but the not conventional "Biomechanical Sense" shows us different "deeper" path to truth.
In fact, to do that, it is essential to deep into the biomechanical tools applied in the throwing techniques movement (Couple and Lever).
*Couple Group*
Analyzing the Techniques of "Couple of Forces" Group, the *General Action Invariant,* (basic shortening movements of distance into the couple of athletes system) , are directly connected to Tsukuri, Kake Phases and the fantastic information is that, in this case, Kuzushi (unbalance) is not a necessary condition, and theoretical speaking it is possible to perform these techniques without unbalance.
 In fact it could be not present in some agonistic applications of couple techniques.

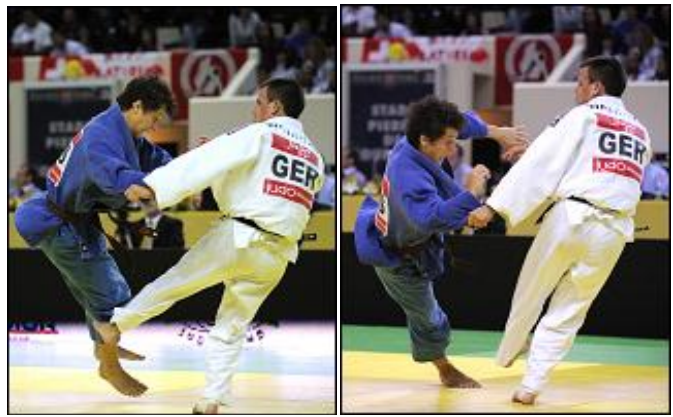

*Fig 7,8  General Action Invariant flowing in Kake phase without Kuzushi (competitive Okuri Ashi  Harai);*

 This means that in such group there are necessary only Tsukuri-Kake phases, in fact the presence of Kuzushi helps and eases obviously the throws but it is not absolutely necessary. Because biomechanically speaking: **Uke is in unstable equilibrium, and the rotation around his center of mass with the fall down is helped by the external gravity force.** Other valid information singled out is that, these more simple techniques, biomechanically speaking, can be applied in competition whatever shifting velocity the couple of athletes system could have. Couple techniques could be grouped in function to their three shortening distance movement called *action invariant* [28], considering the Tori three body's symmetry planes in which lie the couple of forces it is possible to find the inverse movement or throw.

 **A) Sagittal Symmetry Plane** Application of
 1$^{st}$  *General Action Invariants*:  O Soto Gari, reverse direction Mae Ushiro Uchi Mata,
2$^{nd}$ *General Action Invariants:* Harai Goshi inverse application Ushiro Hiza Ura Nage,
3$^{nd}$ *General Action Invariants:* Uchi Mata, inverse (Back) Ushiro Uchi Mata

**B) Transverse Symmetry Plane** Application of:
3$^{nd}$ *General Action Invariants:* O Uchi Gari, Ko Uchi Gari (There are not inverse but opposite from the other side.

**C) Frontal Symmetry Plane** Application of:
1$^{st}$ *General Action Invariants:* Okuri Ashi Harai; opposite Okuri Ashi Harai from the left
3$^{nd}$ *General Action Invariants:* Ko Soto Gari opposite Ko Soto Gari left side

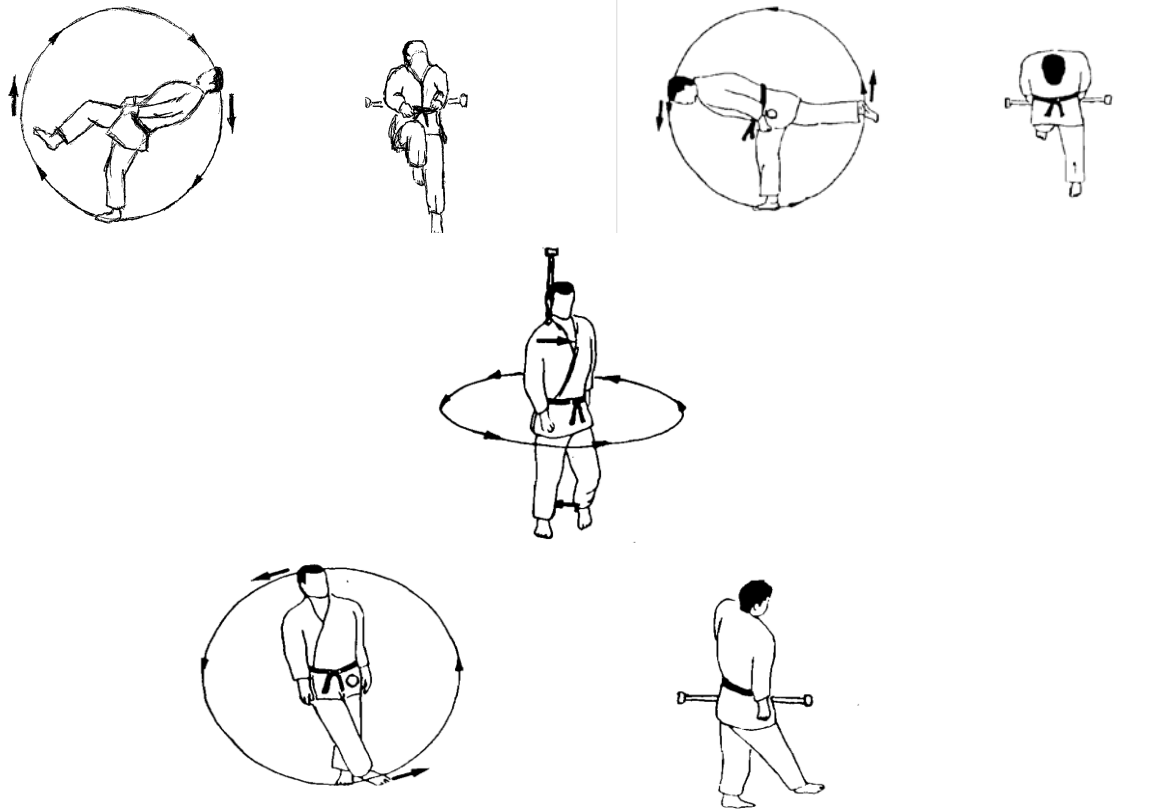

*Fig9-12. Couple in the three planes of Body symmetry: Basic Movements
A) Sagittal, B) Transverse C) Frontal [29, 30]*

To test that kuzushi is not necessary for this group **it is possible to throw someone**, in still position o slow moving as teaching situation**, by De Ashi Barai or by Uchi Mata (**for example**) without grips and then without unbalance.** (True but out the common sense!) . Because grips are responsible, among other function, of the application on the Uke's body of the unbalancing forces, this notation underlines that theoretically speaking Couple Throws are the main eligible throwing techniques for a *Limit Judo Without grips*

*Lever Group*
From the other side Lever throws are heavily dependent from kuzushi, this means that they depend strictly from grips and their unbalancing action.
In other words Lever group techniques are not eligible for a *Limit Judo Without grips.*
 In general considering the degrees of freedom connected to the superior kinetic chains, related to body movement, it is easily understood that all the potential movement in the *Specific Action Invariants,* for the superior chain, in the Kuzushi actions are connected to the three degree of freedoms of the Achromium Joint (shoulder).

From the other side the most important part of the inferior actions are in charge of two Joints hip and knee, less part to the system foot/ankle ( the first for setting better his ones' body in relation to Uke' s body, and the second ones to applying only fulcrum in the lever ).
Biomechanical analysis assure us that Kake phase, for the techniques of physical lever, is the result of the interconnected work performed by both kinetic chains in different time steps.
1) At first the superior chain starts unbalance and open space for the body into adversary grips,
2) then the *General Action* (distance shortening) is developed
3) Followed harmonically by the connected work of both *Inferior* and *Superior Action Invariants* through the abdominals and trunk muscles.

These techniques need more skill in harmonic chains related movements, than couple techniques; in fact often they are ineffective because lack in harmony into one of the previous movements is able to stop the throwing result.

In term of biomechanics for example the same *General Action Invariant* ( **full rotation**) as shortening distance movement ( considering only the inferior chains) splits by three different *Inferior chain action* from up to down, into three well known techniques Seoi nage standing, Seoi Otoshi and Suwari Seoi.

All Kuzushi-Tsukuri movements flow into Kake phase of physical lever application with variable arm, from the most energy wasting one to the lesser ones.( See fig. 13,14,15)

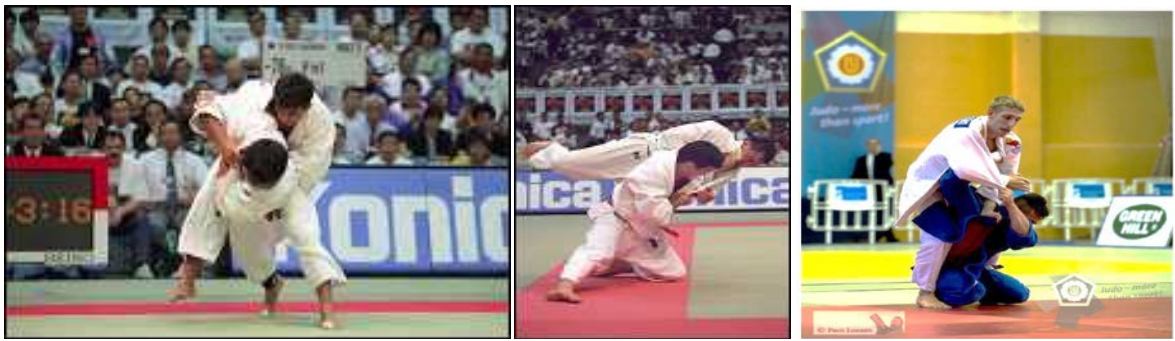

*Fig 13,14,15 Inferior Chain Action Invariant (Seoi); Inferior Chain Action Invariant (Seoi Otoshi) (Finch); Inferior Chain Action Invariant (Suwari Seoi)*

The presence of interconnected movements makes this techniques biomechanically speaking more complex, furthermore on the basis of the physical principle involved for their useful application whatever shifting velocity the couple of athletes has in competition, these techniques need to be applied a stopping time (very short).

**For these techniques if Kuzushi is essential, grips are essential** (Common sense).
But if we speak in term of Biomechanics it is interesting, remembering the Galilean relativity to think that grips are for these techniques a "necessary" approach to interaction.
This means that contact, connecting the two bodies in one whole unit, is essential to apply throwing forces, but **connection** could be accomplished not only by two, but also by one side only. In other terms, the first things that grips do is to connect two athletes in a Couple System changing their.
If somebody grips the opponent and the opponents does not grip him, they are anyway connected in one System of Couple of Athletes.
Connection is assured by contact point, not only by our hands but also by opponent's hands alone. Then from this point of view**:**

**if grip is accomplished by opponent's one hand, the two bodies are "in such way" connected, and for the athlete connected without taking grip it is possible, utilizing the opponent's connection strengthened by his hands on the gripping arm,** (*if the grip is not strengthened opponent could open hand and solve connection*) **to apply a throwing techniques using his own body's weight (internal Tai Sabaki lowering the centre of mass) as unbalancing and throwing force in the direction of gripping arm.**

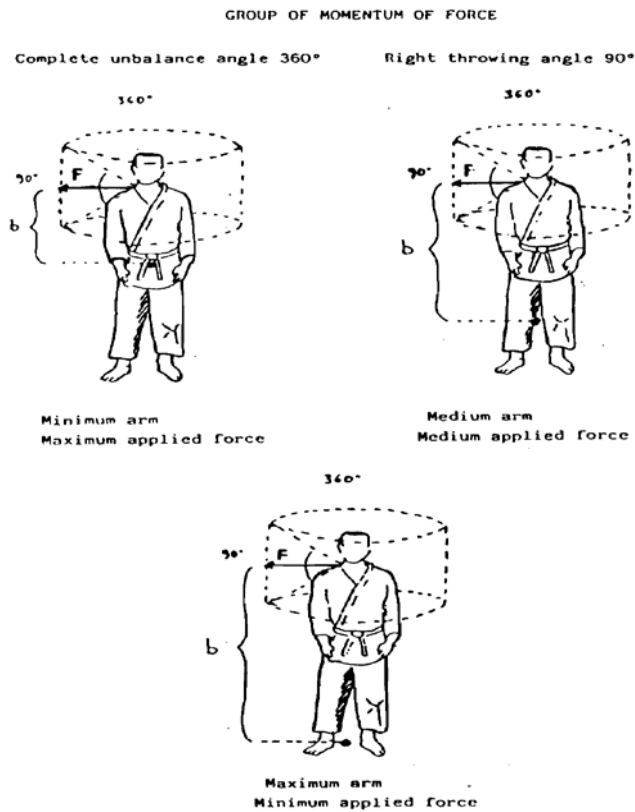

*Fig 16 Complete and right angles to unbalance and throw Uke with physical lever.[29,30]*

Resuming our knowledge and our biomechanical findings, grips are tools with a lot of roles, we list the most important among the many roles, that grips can undertake in high level competitions:

1) ***Connecting role***: at first they connect the two athletes in a one system: the Couple of Athletes System.
2) ***Driving role***: they by the push/pull activity let drive the opponent in a potential favorable position.
3) ***Stimulating role***: they by opportune push/pull actions try to produce good opportunities by previewed reactions
4) ***Shortening role*** : they by right actions shorten the distance between Athletes letting useful tsukuri positioning
5) ***Alerting role***: they can be used to receive from the adversary body's movements alert about his attack action
6) ***Advising role***: they can also receive information from the adversary's body about his movements
7) ***Mastering role*** : they master the distance into the Couple of Athletes system
8) ***Slowing down role*** : they are able with the body weight ( Japanese tool) or with the curled body's position ( Russian Tool) to slow down the opponents' speed movement

9) *Creating role* : they can create Innovative or Chaotic form of throws transferring forces in nontraditional or non rational directions by arms ( Superior Specific Action Invariant )
10) *Avoiding role*: to avoid the adversary's attack technique utilizing his impetus to throw him (ex. Uchi Mata Sukashi)
11) *Destroying role:* to destroy the opponent tentative to build up an attack technique
12) *Active role*: to transfer to the competitor's body an impulse to realize throwing techniques by arms (Superior Specific Action Invariant).
13) *Passive role*: to stop the impetus and the movement of the competitor during his throwing technique.

But the lesser analyzed role is the connecting role, and on this aspect it is possible to study and develop the capability to apply "*judo without grips*." Because the throwing judo techniques belong to the two classes: Couple and Lever, and since the Couple techniques are friction independent, whereas Lever techniques are friction dependent , then based on the unbalance action, it is necessary to deeper also the mechanical results of judo unbalance action. Remembering that human body standing is in a position of instable equilibrium, unbalance:

1. Makes impossible motion, when applied ( the unbalanced body is freeze up in his motion capabilities),
2. When it is applied in classical Kano didactic way , (the unbalance force direction more or less 45° up) athletes' body become also rigid making easier the throwing action,
3. When it is applied in real competition, the classical unbalance concept changes into *"Breaking Symmetry"* action [7] that ends by a collision between athletes' bodies "Butsukari" (ぶつかり) in Japanese.
4. When applied in rotational way both classical unbalance and "Breaking Symmetry" action are very difficult to avoid.

From the previous line on unbalance and Breaking Symmetry, it is possible to infer that the more effective way to apply Lever techniques in Limit "no grip" situation, in real competition is the rotational application.

## 5. At limit of practice : "Astonishing Judo without grips"
### Basic theoretical approach

In judo match analysis as well known and analyzed, there are four situations that can be identified not only as interval time during the competition evolution but also as attack space separating the opponents and subsequently by kind of grips that fighters can apply.
 I. Athletes separate in no grip situations.
 II. One handed grip.
 III. Classic two handed grips.
 IV. Curled up grips.

These situations are particular moments of the most rational approach referring to "Competition Invariants".

The first situation today, with the last refereeing rules, is the area specialized of grips fight. Grip fighting, defined in previously as "arms wrestling" is a wide field of deep study among National Federation, but in high level competitions it is also, more often, the most boring part ones.

Dr. Kano underlined the concept of "Ju" in the use of the opponent's force, and in the text it is shown a more subtle way to overcome this boring phase with the application in a more general meaning of "Sen No Sen" concept.
This special way is, often, applied to this specific situation as very effective and useful tool by some Korean champions, today also applied by European Athletes like French or Dutch competitors.
The wide area of application of judo without grips is obviously the transition period between no grips and stabilized grips, in other words, every time after matte application.
The basic theoretical approach is grounded on the meaning of connection and the physics of two bodies.
The theoretical physical system is organized as two masses connected by a spring as whole system (Couple of Athletes) in two cases:
- when one athlete grips the other with one arm
- when both athletes are gripped together

If we consider Interaction (throwing action) then we must analyze an articulated system of two bodies moving into the three symmetry planes Sagittal, Transversal, Frontal .
As principle, the best way to make sure to score is the throwing action applied with an internal rotation respect the adversary's arm gripping; in such way , blocking the gripping arm, Tori assures himself that Uke can't use this arm to stop the throwing action, (normally for Tori these movements are as example: for Uke's right Tsurite arm, clockwise and for Uke's left Tsurite arm counterclockwise ).

## 6. *Physical and Biomechanical Reassessment*

Finding nontraditional way to make judo, we try to find by biomechanics "original" way to apply physics in high transitory non standard situations.
In classical Newtonian physics the found nontraditional ways are labeled in these items:

For the *lever group techniques*
   a. Connection between two bodies.
   b. Dynamics of a body in a field of elastic forces
   c. Clockwise and counterclockwise rotation
   d. Quasi-Plastic collision of extended bodies
   e. Full rotation with free fall

For the *couple group techniques*
   a'. Direct shortening distance
   b'. Couple application in Frontal plane
   c'. Couple applied in Sagittal plane
   d'. Final Vortex Application with free fall

*a. Connection between two bodies*

This is a very simple concept, but breaking the judo Common sense, **To Connect** in judo sense means to link two athletes' bodies by their grips, but if we think that **only one arm that links with a grip the two bodies is a real bridge between them,** this means to connect, then in such case, if the gripped athlete prevents the adversary's grip detachment blocking the arm that grips him, he can throw the adversary without take an usual judo grip.
In this way the unbalance force is transmitted to the adversary by his bridging arm, this way that we claim *"judo without grips"* could be also called Sen No Sen on grips, was applied mainly by Korean Athletes years ago, today in many different variation but with the same principle it is a common knowledge of many skilled athletes in the world, thanks to match analysis.

*b. c. Dynamics of a body in a field of elastic forces with clockwise and counterclockwise rotation*

The two Bodies that play judo are mainly connected by arms that apply push pull forces in every direction. In this sense we can compare the body motion at a situation of dynamic of a body in a field of elastic internal forces.
The Bertrand theorem assure us that the only closed orbit for the elastic generalized force

$$F = -kx^\alpha$$

There are only for α = 1; or α = -2 as already demonstrate in [31].
Interesting it is to analyze the trajectories that Tori normally use to apply lever techniques in this situation. How it easy to see in the next figure all trajectories are similar to it, because Tori taking firmly a contact point performs a fast rotation to apply the lever and throw Uke.
It is interesting that always speaking in term of elastic field the class of trajectories applied by Tori are similar to a capture trajectory of a particle by another in elastic field with internal forces

$$F_{ab} = -F_{ba} = F$$

In this case the two particles act like the two athletes bodies during the throw. [32]

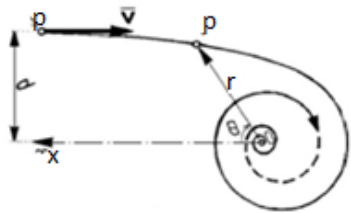

*Fig .17 Trajectory of a capture particle similar to Tori's throwing trajectory in judo interaction*

### d. Quasi-Plastic collision of extended bodies

The end of the previous shown trajectory flows into a projection by a lever techniques that starts with a collision, which can be considered a quasi plastic collision because the two athletes are strictly connected together, but obviously their bodies are not merged, then it is a collision quasi-plastic of two extended bodies.
In this case considering both athletes of more or less equal mass, the equations are very easy to obtain. [33]
If they have different starting velocities, after the contact they move connected till to the fall down that can be considered as a free fall.
In this case considering always negligible the gravity force, that increases greater and greater his importance into the motion after the collision, till to landing of Uke body ( free fall); it is possible to write for the early instants of the collision, remembering that is a rotational impact: [34]

$$mv_1 + mv_2 = 2mv$$
$$\text{the conservation of angular moment give us } I_1\omega_1 = (I_1 + I_2)\omega_f$$
$$or \quad \frac{\omega_f}{\omega_1} = \frac{I_1}{(I_1 + I_2)}$$
$$\text{the impact is totally inelastic, and the loss of kinetic Energy is:}$$
$$\Delta K = \frac{1}{2}I_1\omega_1^2 - \frac{1}{2}(I_1 + I_2)\omega_f^2 = \frac{1}{2}\frac{I_1 I_2}{I_1 + I_2}\omega_1^2$$

### e. Full rotation with free fall

Complex rotational application of judo throwing techniques that will be analyzed in this section are connected to the tactics of first contact, applying a lever techniques with a fast and complete rotation, like spinning top at variable mass or vortex
It is, for Tori as observer, a clear study of Forces applied in a rotating reference frame.
At first it is important to evaluate the velocity transformation formula from the inertial to the rotating frame, this means how the speed is evaluated by people (public as observer) and Tori as observer during the execution of a rotating throws, as already demonstrate in a previous paper [6] the result is:

$$v = V + \left(\frac{dr'}{dt}\right)_O = V + (v' + \omega \wedge r')$$

It is important to evaluate the general equation of motion of this variable mass spinning up, remembering the classical Newton approach to the rotational dynamics, [35] we can write:
The torque on the first athlete $\tau$ will be
$$\tau = r\,F$$

Where $r$ is the radius between the centre of mass and the point where the force $F$ is applied, in term of rotational dynamics this equation can be written also as:

$\tau = I\dfrac{d\omega}{dt} = mr^2 \dfrac{d\omega}{dt}$  it is very easy to solve this equation if the athlete is up-righted like a symmetric spinning top, because as long as the torque is applied in such a way as to increase (or decrease ) its rotational speed around the ẑ-axis, this is just a one-dimensional equation and offers no surprises. [36]

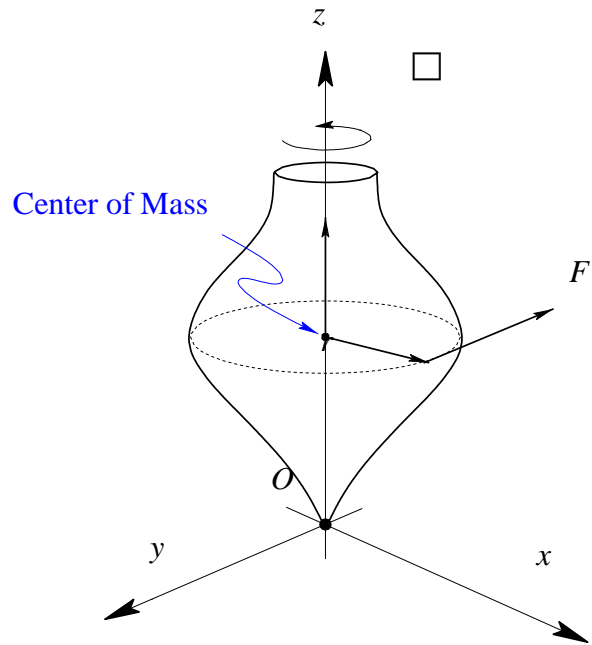

**Fig. 18  Spinning Top**

It is necessary to remember that if we use the Euler representation of a rotating rigid body they produce a non linear system of equation not always resolvable.

However interesting are the equations of motion of the Euler angle found by Garanin [37], that are shown in the next:

$$\dot{\theta} = \left(\dfrac{1}{I_1} - \dfrac{1}{I_2}\right) L \sin\theta \sin\psi \cos\psi$$

$$\dot{\varphi} = \left(\dfrac{\sin^2\psi}{I_1} + \dfrac{\cos^2\psi}{I_1}\right) L$$

$$\dot{\psi} = \left(\dfrac{1}{I_3} - \dfrac{\sin^2\psi}{I_1} - \dfrac{\cos^2\psi}{I_1}\right) L \cos\theta$$

Note that equations for $\dot{\theta}$ (nutation) and $\dot{\psi}$ (spin) form an autonomous system of equations that can be solved as first step, after that the equation for the precession $\dot{\varphi}$ can be integrated using the previous found spin obtaining the demanded solution-
If, however, athlete *tilts* his rotational axis through an angle $\varphi$, as shown in Figure, the situation gets a little more complicated and a lot more interesting respect to a spinning top. In the case of the tilted athlete shown in Figure , gravity pulls down on the centre of mass of the athlete, which would pull a non-spinning athlete (because it is in unstable equilibrium) downward and simply increase the tilt angle φ as the athlete falls down. Normally in a spinning top the torque, and thus the change in the angular-momentum vector, is perpendicular to the axis û, which leads the top to move "sideways" in a circle around the *z*-axis, and this motion is called *precession*, but this well known phenomenon is nullified in our case by the increased mass of the system that after the (quasi-plastic collision) firmly connect together the two athletes bodies.

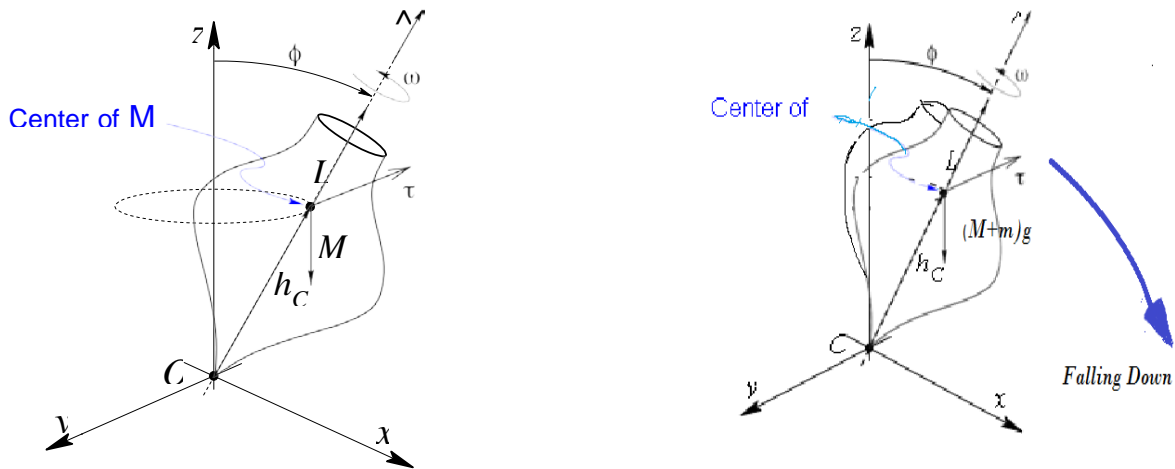

*Fig.19, 20 Tilting Spinning Top; Variable Mass Falling Spinning Top.*

The three - dimensional equation becomes:

$$\tau = r \wedge F$$

$$\tau = \frac{d(I\omega)}{dt} = \frac{d[(M+m)\omega]}{dt}$$

Abruptly after contact, mass more or less double, system velocity drops down and external gravity force overcoming the potential precession motion helps the bodies to fall down. [38]

### a'. Direct shortening distance

For Couple Techniques (more biomechanically simple) see a previous paper [28] the first part of skill action is to short the distance between athletes' bodies and after to apply a Couple to the adversary' body.
In the first contact phase the only fastest and useful way to short the distance contacting the adversary' body is the right line shortening.

### b'. Couple application in frontal plane

The application of a Couple that produce the rotation in the frontal plane, with the actual refereeing rules, is carried out essentially by legs.

### c'. Couple applied in Sagittal plane

Also the application of a Couple that produce the rotation in the sagittal plane, with the actual refereeing rules, is carried out essentially by legs, but there is one notation, if the Uke body twists versus his back only the Sagittal Couple is sufficient, if Uke body twists forwards Tori must apply a vortex action (a torque that turns Uke on his back) to obtain Ippon.

### d'. Final Vortex Application with free fall

The well known equation of rotation of a body is :

$$L = \frac{d(I\omega)}{dt}$$

With this little trick Tori would be able to obtain Ippon, following the well known refereeing rules, Uke back on the mat (Tatami). [39]

## 7. Technical Principles of Jūdō Revised

All the previous discussion let us able to revise the technical judo principles that normally are applied and valid during the high level competitions.
In effect during the transitional phase of **the first contact** that can happen time and time again during a competition, the grip rules can be changed.
It is not necessary to grips with a known kumi kata (classic or not) the adversary, but it is sufficient to establish a contact ( in a broad sense) with the adversary body, if that happens One of the two athletes is able to apply a lever throwing technique.
From the other side always during **the first contact** time, if the athletes are specialized in Couple throws, they need only to shorten the distance to apply the couple on the adversary body.
The biomechanical analysis assure that the body right or left rotation to apply Lever techniques is the common denominator for these throws linked to a quasi plastic collision and a downward leap.
Some time also the fast and fully rotational application of these techniques are allowed utilizing as contact wrapping their arms around the adversary's body.
For the Couple techniques easier the skill action are only straight shortening distance and Couple application

## 8. Practical application in high level competition

All the tactics previous described are always applied in high level competitions as sen no sen on adversaries' grips.
In effect the more subtle way to apply Sen No Sen, in such situation, is to throw **without grip,** using the opponent's grip that let him to stay connected.
Very clever expert of this approach was Won Hee Lee from Korea who applied this method many times.

Lee, timing, makes the right space with internal Tai Sabaki ( **General Action Invariant** based on Hando No Kuzushi), stabilizes his contact, using the well known: two hands against one arm grip to strengthen the connection, and throws the opponents in harmony with the two hands contact as shown in the next Sequences.

It is not matter of techniques everybody can apply his preferred technique in the specific situation (compulsory is the Lever Group belongings).

The internal Tai Sabaki more often follows the opponents pull, and with it athlete shortens the distance and set his body for throwing action, if athlete performs external Tai Sabaki, he avoids opponents action ( wasting some time) and can apply some couple group throws like Innovative Uchi Mata. Avoiding and Sen No Sen form strategies, depend also from weight and personal speed and skill of athletes.

An opportune and ad hoc training of these situations would be useful to apply, carefully, such "Trick" biomechanically grounded both on the Galilean Relativity and a better understanding of grip concept in real competition.

Application of Lever Group Tactic at first contact

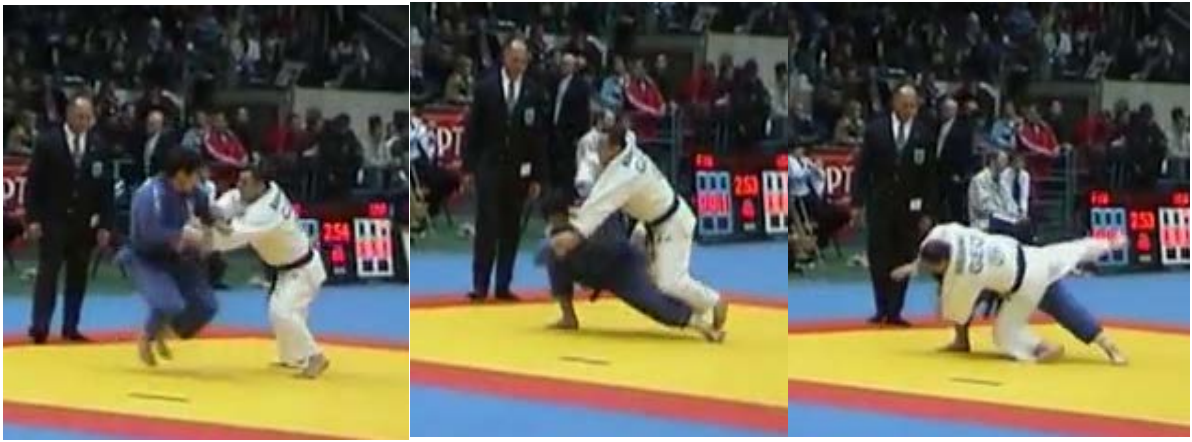

*Fig 21-23 White grips straight left hand, Lee Attacks two against one hand left Tai Otoshi*

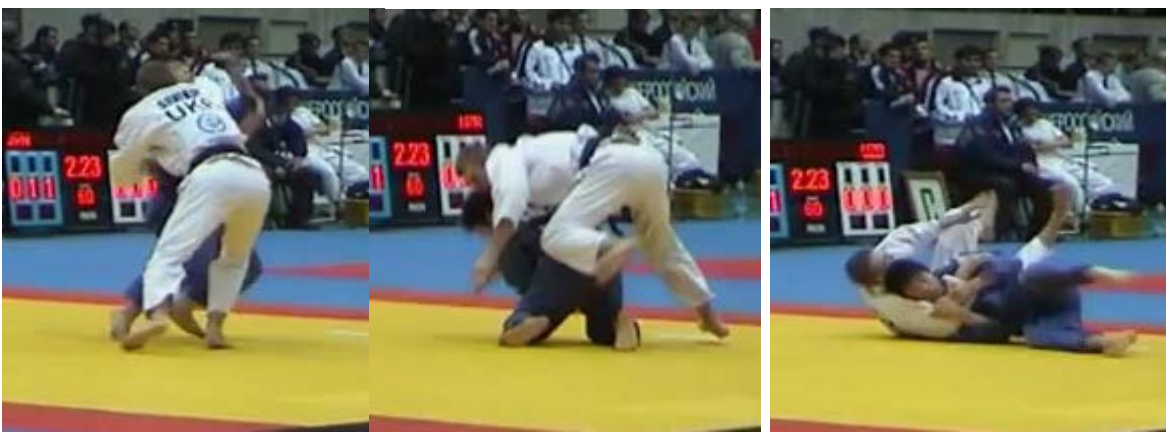

*Fig 24-26 White grips right from top, Lee attacks two against one hand right Suwari Seoi*

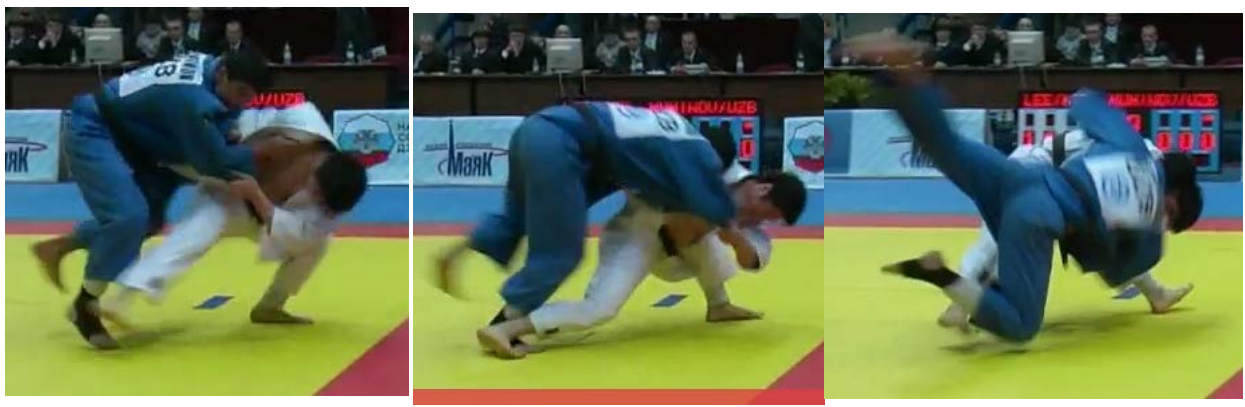

*Fig.27-29 Blue grips right Lee's left side, Lee attacks two against one hand right Tai Otoshi*

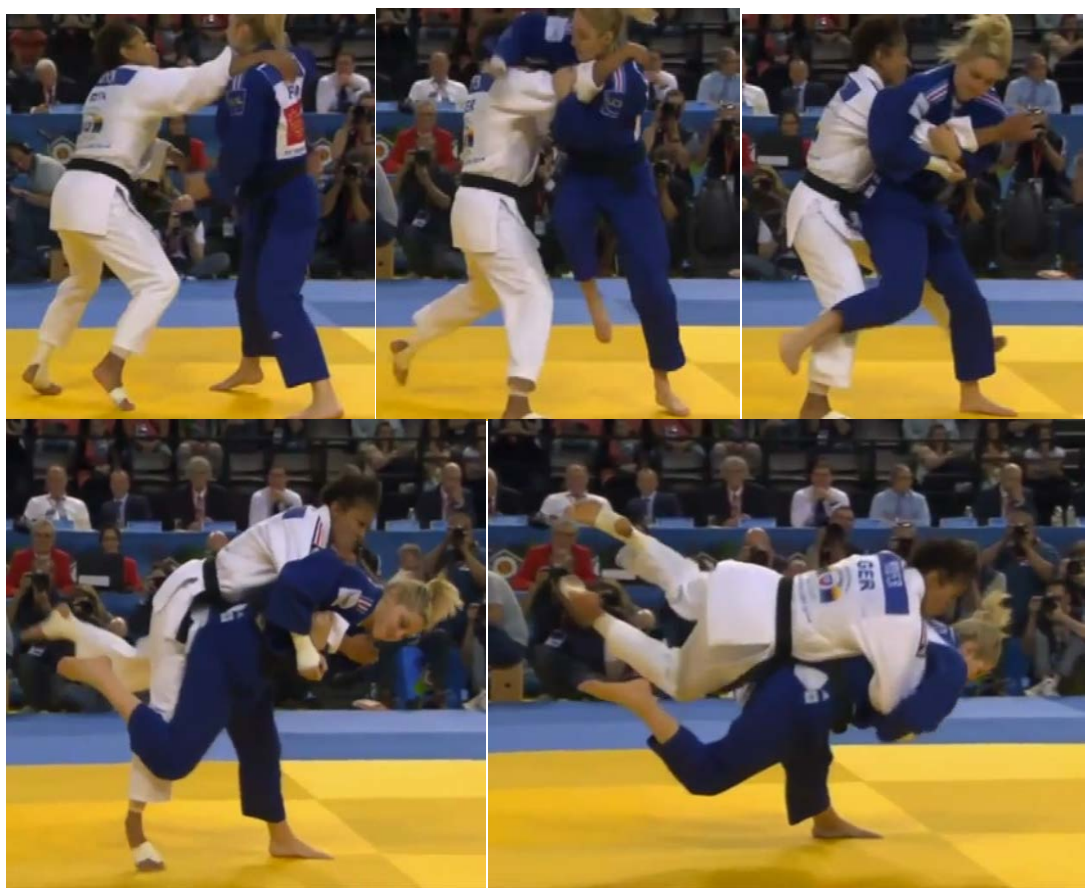

*Fig. 30-34 Tactical application of O Soto Makikomi at First contact, blocking the one Tsurite arm contact*

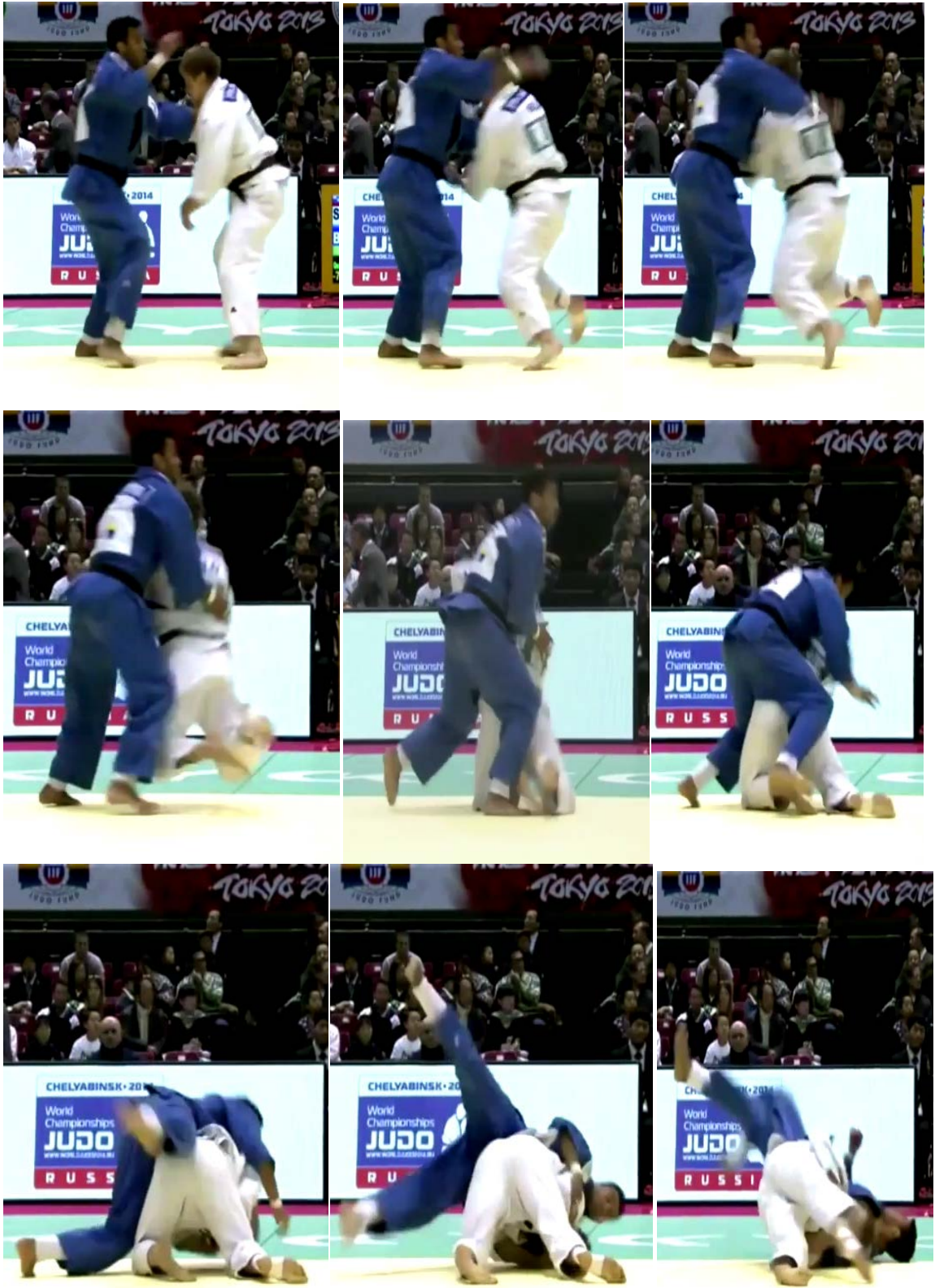

*Fig.35-43 Oblique Suwari Seoi tactic applied at First Contact, gripping the left Tsurite adversary's arm.*

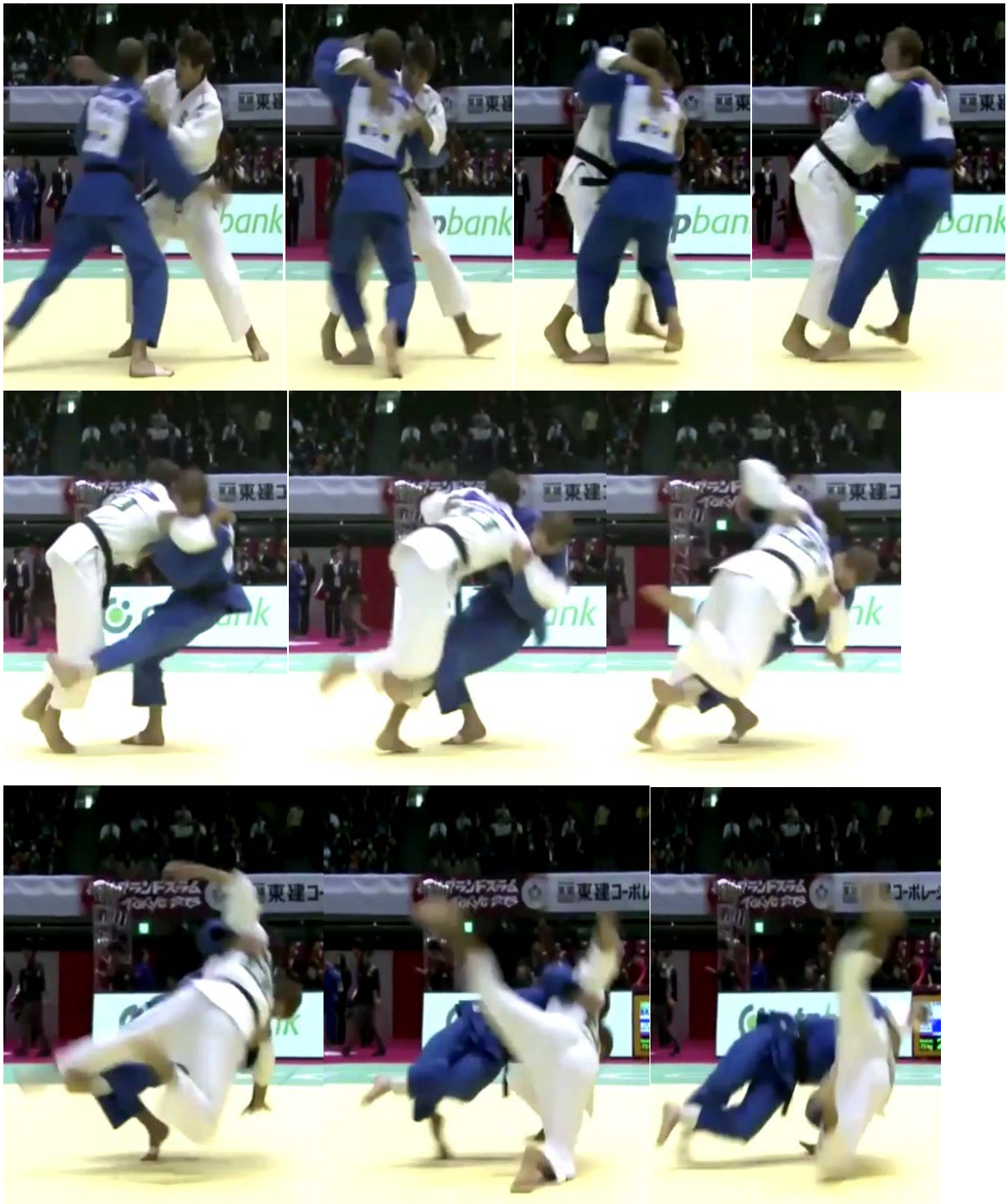

*Fig 44-53 Full rotational application tactic at first contact Rotational Hiza Guruma*

**Application of Couple Group Tactic at first contact**

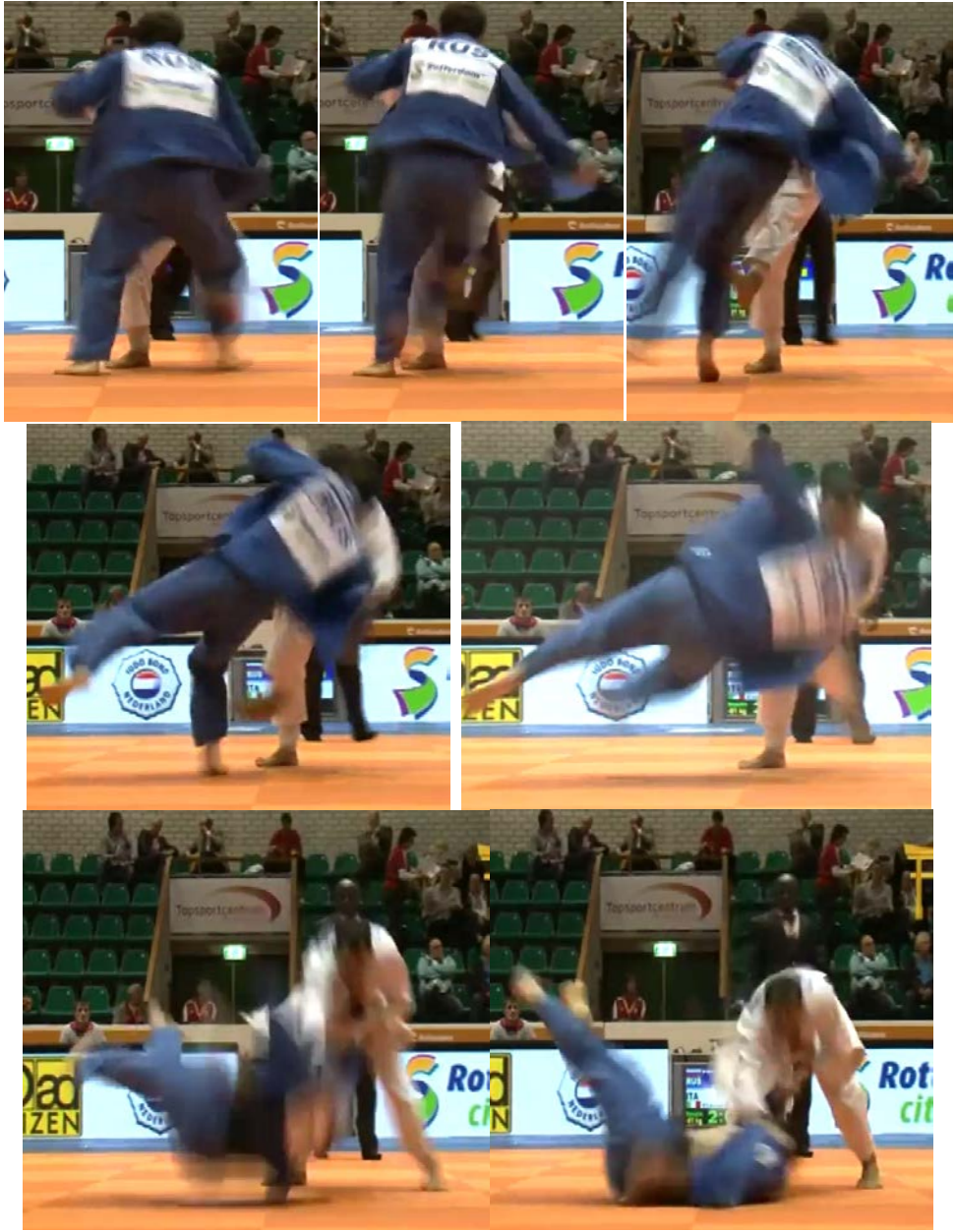

*Fig 54-60. Shortening forward distance and Couple application in the Frontal Plane at First Contact*

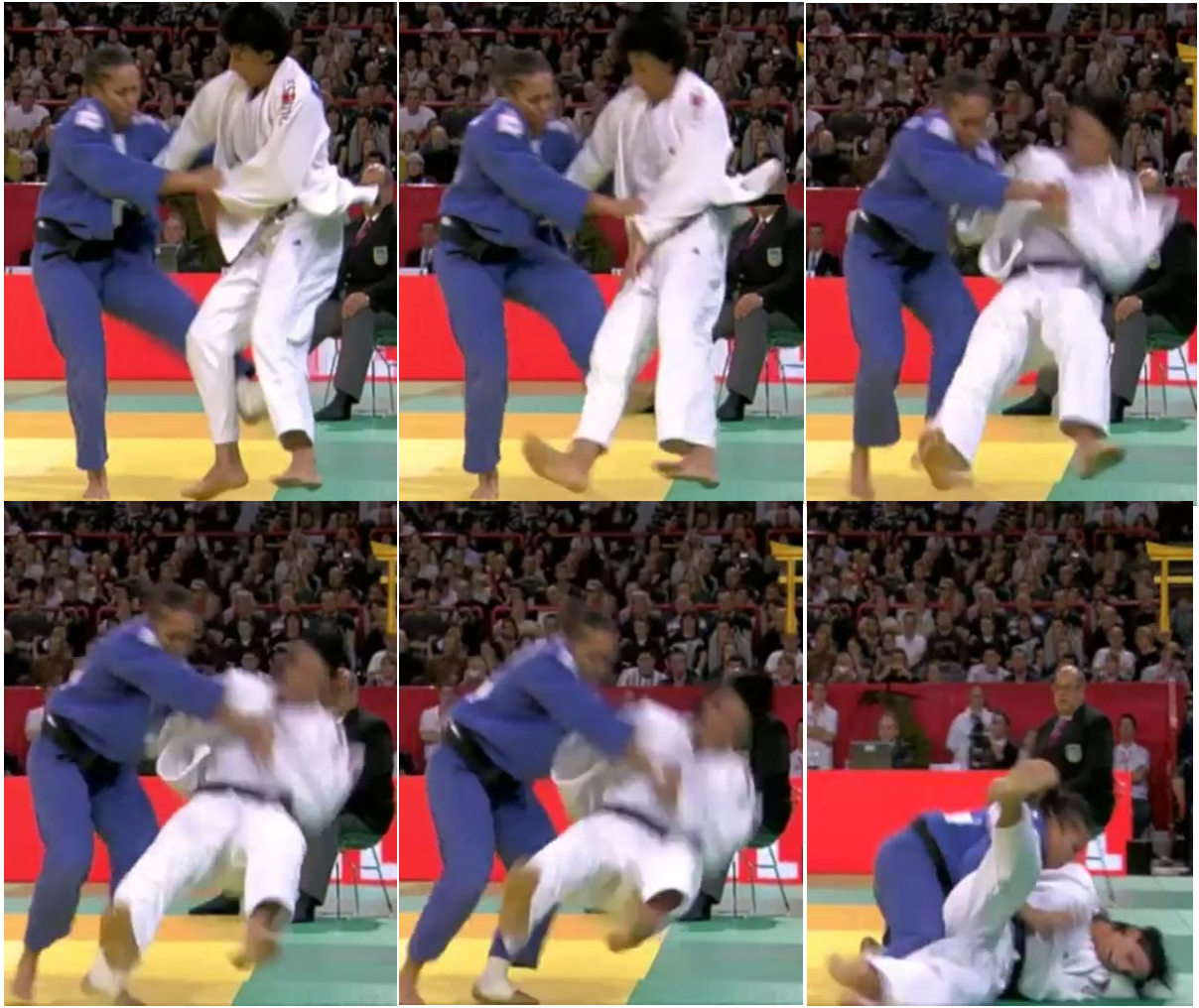

*Fig 61-66. Distance Shortening and Couple applied in a frontal plane at First Contact*

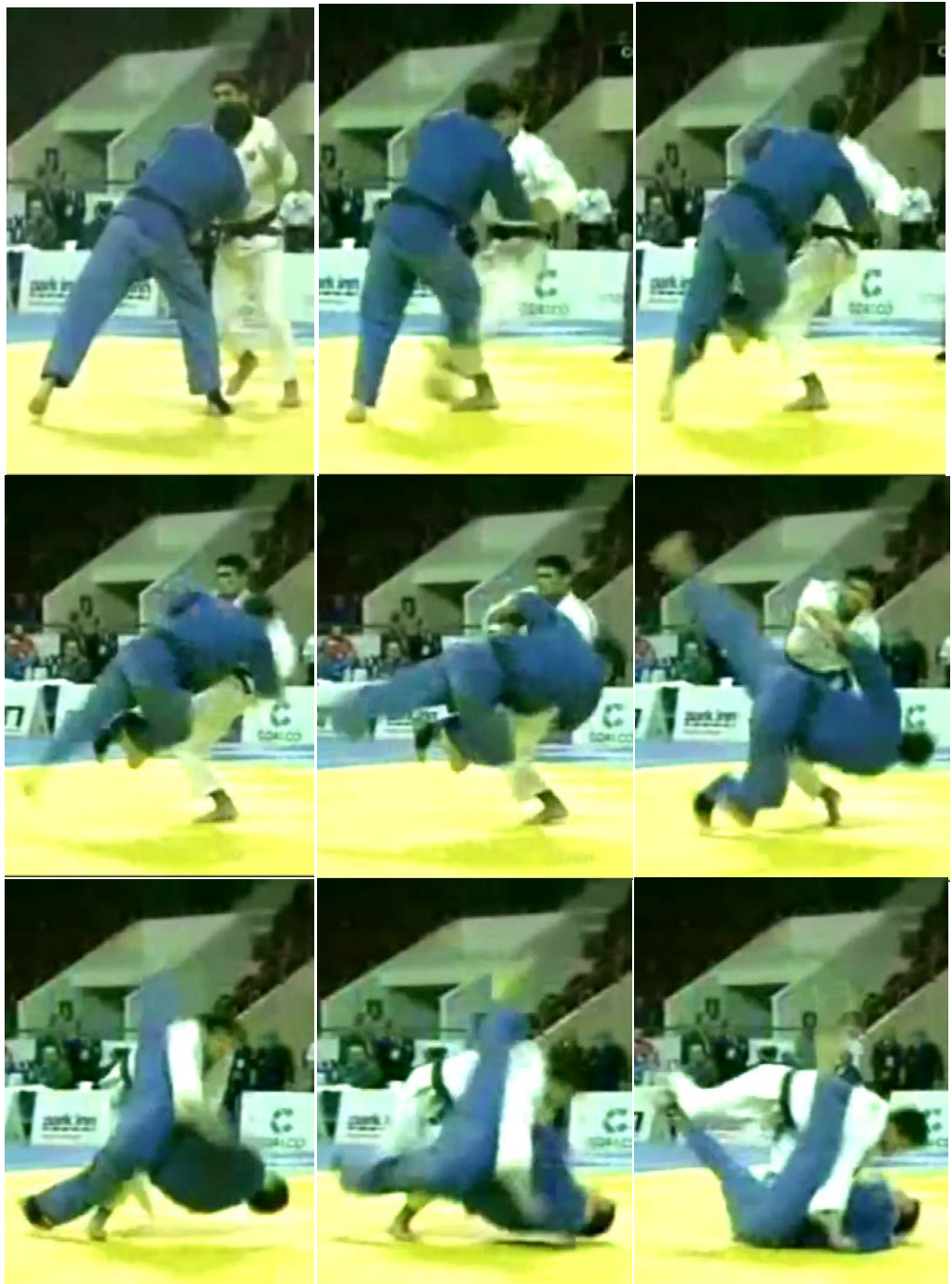

*Fig 67-75. Couple Application in the frontal Plane at First Contact*

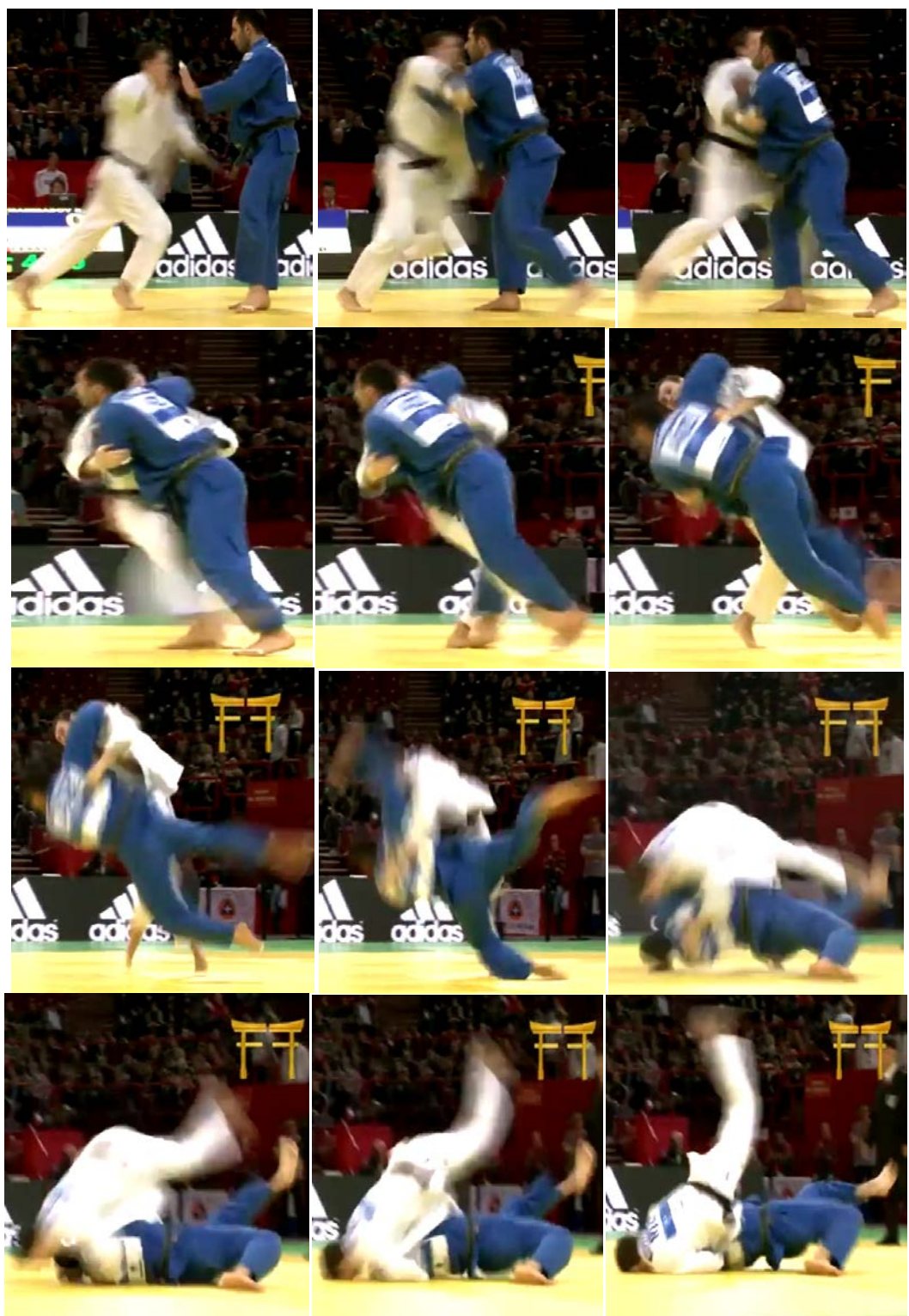

*Fig 76-87. Shortening distance and Application of Couple forward in sagittal plane with clockwise rotation at First Contact*

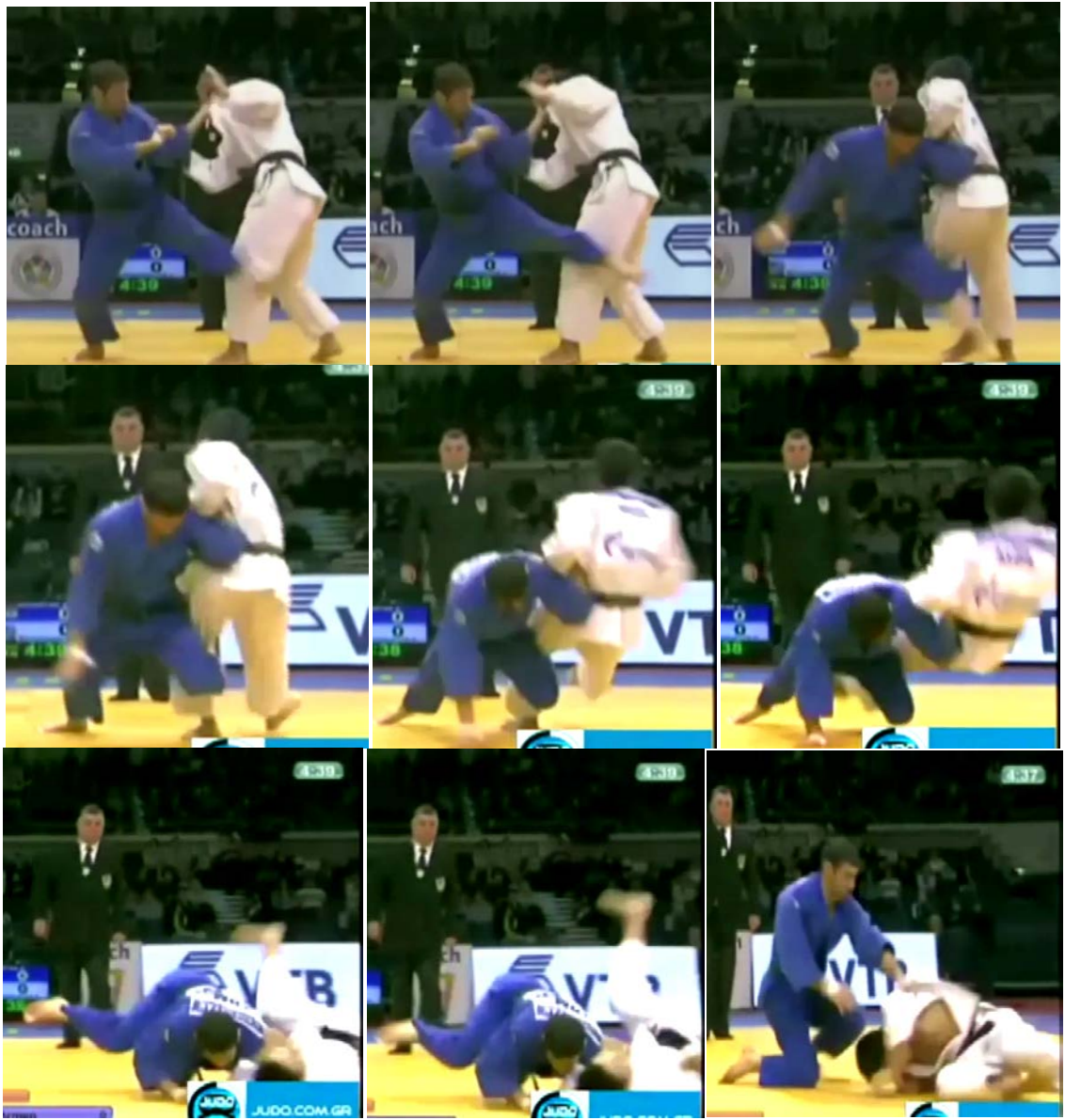

*Fig 88-96. Shortening distance and Application of Couple backward in sagittal plane at First Contact*

## 9. *Conclusion*

In this paper was performed the identification and a biomechanical evaluation of tactical tools that can be utilized at first contact in high level competition.
This is the third paper focalized on finding for high level competition tools and methods useful to enhance judo throwing techniques effectiveness.
The first paper was committed to the direct attack enhancement; [6] the second to the tools applied for combination and action reaction attack enhancement [7].
Now the study of first contact situation is performed, and as soon as possible will follow the identification of biomechanical tools for enhancement of counters.
In this first contact study, biomechanics was able to single out the means and practical tools that Tori can apply to obtain high technical vantage or the victory in high level competition.
Different tools are applied for the Lever group throwing techniques and for the Couple group.
By the knowledge of these tools it is possible to train athletes in tactics, or to enhance their judo effectiveness, often breaking the old common sense that lingers in our world

For the ***lever group throws,*** remembering that these techniques are more complex in term of movements and body coordination the practical steps utilized could be divided in five terms following the Physical model utilized:

- A. Connection between two bodies.
- B. Dynamics of a body in a field of elastic forces
- C. Clockwise and counterclockwise rotation
- D. Quasi-Plastic collision of extended bodies
- E. Full rotation with free fall

For the ***couple group throws,*** depending from their biomechanical relative simplicity the practical steps are following the Physical model utilized are only two or three:

- a'. Direct shortening distance
- b'. Couple application in Frontal plane

- a. Direct shortening distance
- b. Couple applied in Sagittal plane ( Uke backward direction)

- c. Direct shortening distance
- d. Couple applied in Sagittal plane (Uke forward direction)
- e. Final Vortex Application with free fall

The only two plane preferred as main tactics, are the planes in which it is possible to apply a throw with minimum of energy consumption and also less movements, that means fastest application , against Uke reaction time.
The astonishing results obtained by biomechanical analysis, compel us to revise well stabilized concepts in judo that are utilized often without any criticism.
First contact is a very important part of tactics studies for competition based on high attack speed, non conventional grips, and use of time in high level competition.
Remembering that first contact starts after every, matte –ajime , it is easy to understand that this phase happens many times during competition and its importance is very high in the strategic and tactics preparation.

Obviously a critic analysis, of the tools found, shows that they are successful because are grounded sometime on Ukes' silly mistakes in movement or in grasping the grip.

Some other time these tools are effective because Tori uses non conventional way to attack or the concept of contact-connection instead of classical grips.

For the lever group, for example, more often the tactical trick is applied on a mistaken utilization of Uke tsurite arm.

From the previous analysis some different utilization or application of judo principles flows, but these principles already identified and explained could be applied specially during the special situation of *"First Contact"*.

## 10. References

**[1]** Kawamura T, Daigo T, editors.: ***Kōdōkan New Japanese-English Dictionary of Judo*** (和英対照柔道用語小辞典). Tōkyō: Kōdōkan Institute; 2000.

**[2]** Cortell Tormo, Pérez Turpin, Lucas Cuevas ***Handgrip strength and hand dimensions in high-level inter-university judoists*** Archives of Budo. Volume 9, 2013, N°1

**[3]** Bonitch-Góngora, . Bonitch-Domínguez, Feriche, Belén ***Maximal isometric handgrip strength and endurance differences between elite and non-elite young judo athletes*** 2013

**[4]** Bonitch-Gongora, Bonitch-Dominguez, Padial, Feriche. ***The Effect of Lactate Concentration on the Handgrip Strengthduring judo bouts*** Journal of Strength and Conditioning research Volume 26 July 2012 N° 7

**[5]** Detanico, Arins, Dal Pupo, Dos Santos ***Strength Parameters in Judo Athletes: An Approach Using Hand Dominance and Weight Categories*** Human Movements Vol 13 January 2013 N°4

**[6]** ***Studies on Judo techniques with respect to distribution of body weight***

**[7]** Sacripanti : ***A Biomechanical Reassessment of the Scientific Foundations of Jigorō Kanō's Kōdōkan Jūdō***, Cornell University Library. Available from URL: http://arxiv.org/abs/1206.1135

**[8]** Sacripanti: ***Judo how to enhance tactics in competition: biomechanics of combination and action reaction attacks.*** 2014 Cornell University Library. Available from URL: http://arxiv.org/pdf/1401.1102 .

**[9]** Sacripanti ***Judo how to enhance effectiveness of direct attack throws - Kano's dream : Judo rotational application"*** 2014 Cornell University Library. Available from URL: http://xxx.tau.ac.il/abs/1405.1982v1

**[10]** Idarreta.. ***Estudio de la ambidextría de ejecución técnica en jóvenes judokas de élite españoles.*** 2005 available at URL: http://www.rendimientodeportivo.com/N007/Artic034.htm.

**[11]** Sterkowicz, Sacripanti, Katarzyna Sterkowicz-Przybycień ***A techniques frequently used during London Olympic judo tournaments: a biomechanical approach*** Arch Budo vol 1 2013 N° 4

**[12]** Bocioaca ***Technical and Tactical Optimization Factors in Judo*** Procedia - Social and Behavioral Sciences 117 (2014)

**[13]** Kwon, Cho, Kim. ***A Kinematic Analysis of Uchi-mata (inner thigh reaping throw) by Kumi-kata Types in Judo*** Korean journal of sports biomechanics 2005

**[14]** Kwon, Kim, Cho ***A Kinematics Analysis of Uchi-mata (inner thigh reaping throw) by Kumi-kata Types and Two Different Opponents Height in Judo (2)*** Korean journal of sports biomechanics 2005

**[15]** Kim, Yon ***A kinematic analysis of the attacking-arm-kuzushi motion as to pattern of morote-seoinage in judo*** Korean Journal of Sport Biomechanics Vol13 2003 N°1

**[16]** Kim; Yoon; Kim, ***A Case Study on Kinematical Traits Analysis when Performing of Uchimatia(inner thigh reaping throw) by Posture and Voluntary Resistance Levels(VRL) of Uke in Judo[ I ];*** Korean Journal of Sport Biomechanics Vol14 2004 N°3.

**[17]** Calmet. ***Kumi-kata, distances et rotations de l'attaquant lors des approches et saisies de l'adversaire dans les combats*** 2009 Milano Italy

**[18]** Nobuyoshi, Morio, Mitsuru, Takashi ***The analysis of judo competition in tactics of throw techniques-Comparisons between the men judo athletes and the women judo athletes*** Journal of Health and Sports Science Juntendo University Vol.4 2000 N°3

**[19]** Marcon, Franchini, Jardim , Barros Neto ***Structural analysis of action and time in sports: Judo*** Journal of Quantitative analysis in Sport Vol . 6 2010 N°4 Available at URL: http://www.researchgate.net/publication/227378864_Structural_Analysis_of_Action_and_Time_in_Sports_Judo/file/72e7e529cc3ba6d714.pdf

**[20]** Hernández García, R. y Torres Luque, G. *Análisis temporal del combate de judo en competición*. Revista Internacional de Medicina y Ciencias de la Actividad Física y el Deporte vol. 7 2007 N°25 Available at URL: http://cdeporte.rediris.es/revista/revista25/artjudo46.htm

**[21]** Sterkowicz, Maslei ***An Evaluation of the Technical and Tactical Aspects of Judo Matches at the Seniors Level*** Available at the URL: http://www.judoamerica.com/ijca/sterkowicz/sterkowicz.pdf

**[22]** Miarka, Julio, Boscolo Del Vecchio, Calmet, Francini ***Technique and Tactic in Judo: A Review*** RAMA vol5 2010 N°1

**[23]** Courel, Franchini, Femia, Stankovic, Escobar-Molina ***Effects of kumi-kata grip laterality and throwing side on attack effectiveness and combat result in elite judo athletes*** International Journal of performance Analysis in Sport Volume 14 April 2014 N°1

**[24]** Kajmovic, Husnija; Radjo, Izet. ***A Comparison of Gripping Configuration and Throwing Techniques Efficiency Index in Judo Between Male and Female Judoka During Bosnia and Herzegovina Senior State Championships*** International Journal of performance Analysis in Sport Volume 14 August 2014 N°2

**[25]** Kajmovic, Rađo, Mekic et al. ***Differences in gripping configurations during the execution of throwing techniques between male and female cadets at the European Judo Championship***. Arch Budo vol. 10 2014;

**[26]** Pierantozzi, Nerozzi, Piras, Lubisco ***Analysis of the Fighting Phase Before the First Grip in the Finals of the Judo World Championship 2007*** Journal of Biomechanics Research 2008

**[27]** Piras, Pierantozzi, Squatrito. ***Visual search strategy in judo fighters during the execution of the first grip*** International journal of Sport Science and Coaching Vol 9 February 2014 N° 1

**[28]** Sacripanti: **Biomechanics of Kuzushi-Tsukuri and interaction in Competition.** Paper presented at SPASS International Conference, September 5-9[th], 2010, Lignano Sabbiadoro, Italy. http://arxiv.org/abs/1010.2658

**[29]** Sacripanti A: ***Biomeccanica del Judo.*** Roma: Ed. Mediterranee; 1988 ISBN: 8827203486 9788827203484 [in Italian]

**[30]** Sacripanti A: ***Advances in judo biomechanics research. Modern evolution on ancient roots.*** Saarbrücken, Germany: VDM Verlag Dr. Müller Aktiengesellschaft & Co. KG.; 2010**.** ISBN-13 978-3639105476

**[31]** Goldstein, Poole & Safko ***Classical Mechanics (Third Edition)*** Addison Wesley Pub. 2000

**[32]** Teodorescu ***Mechanical System, Classical models Vol 1,2,3*** Springer 2009

**[33]** Chatterjee *Rigid body collision; some general considerations, new collision laws and some experimental data* PHD thesis 1997
**[34]** Di Benedetto *Classical mechanics* Birkhauser 2011
**[35]** Rose *Elementary Theory of Angular Momentum* John Wiley & sons NY 1957
**[36]** Crabtree *An Elementary treatment of the theory of Spinning Tops and Gyroscopic motion* Longmans Green and Co 1909
**[37]** Garanin *Rotational Motion of Rigid bodies* Cuny graduate Center 28 November 2008
**[38]** Sommerfeld *Mechanics Lecture on theoretical Physics* Academic Press Publisher 1952
**[39]** Sommerfeld *Mechanics of deformable bodies Lecture on theoretical Physics* Academic Press Publisher 1952